\documentclass[a4paper]{article}
\usepackage{graphicx}
\newcommand{\fmfLrt}[3]{\put(#1,#2){\raisebox{-1.0\height}{\makebox[0pt][r]{#3}}}}

\newcommand{\fmfLlb}[3]{\put(#1,#2){\makebox[0pt][l]{#3}}}
\newcommand{\fmfLlt}[3]{\put(#1,#2){\raisebox{-1.0\height}{\makebox[0pt][l]{#3}}}}

\def\fun#1#2{\lower3.6pt\vbox{\baselineskip0pt\lineskip.9pt
\ialign{$\mathsurround=0pt#1\hfil##\hfil$\crcr#2\crcr\sim\crcr}}}

\newcommand{\be}{\begin{eqnarray}}
\newcommand{\ee}{\end{eqnarray}}
\newcommand{\bd}{\begin{displaymath}}
\newcommand{\ed}{\end{displaymath}}
\newcommand{\ba}{\begin{array}}
\newcommand{\ea}{\end{array}}
\newcommand{\bt}{\begin{tabular}}
\newcommand{\et}{\end{tabular}}

\title{\bf
Gell-Mann - Oakes - Renner relation in a magnetic field at finite
temperature.}
\author{
N.O. Agasian\thanks{e-mail: agasian@heron.itep.ru}~~and
I.A. Shushpanov\thanks{e-mail: shushpan@heron.itep.ru}
\\
\\
{\it Institute of Theoretical and Experimental Physics} \\
{\it 117218 Moscow, Russia}}
\date{}

\begin{document}

\maketitle

\begin{abstract}
In the first order of chiral perturbation theory the corrections to
$F_{\pi^0}$ and $M_{\pi^0}$ in a magnetic field at finite
temperature have been found.
It was shown that they are shifted in such a manner that
Gell-Mann - Oakes - Renner relation remains valid under these conditions.
\end{abstract}

\vspace{2cm}


\newpage

1. The investigation of the vacuum properties and hadron structure behavior
under influence of the various external factors is known to be one of the
interesting problems in quantum field theory. At low energies strong
interactions can be adequately described by chiral dynamics of light
pseudoscalar particles - $\pi$-mesons, i.e. chiral effective theory
[1,2]. Besides, in the studying of QCD hadron phase the low-energy relations
at finite temperature [3] in a magnetic field [4]
are very useful.
One of the important phenomenon of low energy physics of pions is
Gell-Mann - Oakes - Renner (GOR) relation [5], establishing
the connection between pion mass $M_\pi$, axial coupling constant
$F_\pi$ and quark condensate $\langle \bar q q\rangle$.
Validity of GOR relation at finite temperature was investigated in [6].
It was found that GOR relation is not affected by low temperature.
Similar consideration of GOR relation in external magnetic field was
done in [7].
In the presence of the field, the axial
$SU_A (2)$ symmetry is broken down to $U_{A}^3 (1)$ corresponding
to chiral rotation of $u-$ and $d-$ quarks  with opposite phases (the
singlet axial symmetry is broken already in the absence of the field due
to the anomaly ). The formation of the condensate breaks down this remnant
$U_{A}^3 (1)$ symmetry spontaneously leading to appearance of a
Goldstone boson, the $\pi^0$-meson.
Meanwhile, charged pions acquire
a gap in the spectrum $\propto \sqrt {eH}$ and are not goldstones
 anymore.

Phase structure of QCD vacuum in a constant magnetic field $H$
at low temperature $T$ was explored in [8]. In the framework of chiral
perturbation theory (ChPT) the dependence of the quark condensate upon $T$ and
$H$ was calculated and showed that shift of the condensate is not a simple
sum of temperature ($\sim T^2/F_\pi^2$) and magnetic ($\sim H/F_\pi^2$)
corrections. Orbital diamagnetism of charged pion gas results in
the additional, "cross" term.
In the present paper we study GOR relation in a magnetic field
at finite temperature.

2. At low temperatures $T\ll T_c$ ($T_c$ is the temperature of the
phase transition with chiral symmetry restoring) and in the region
of the weak magnetic fields \footnote{A transition to the chiral
limit that $M^2_\pi \ll H$. It was shown in [4] that the parameter
of the ChPT expansion in a magnetic field is $\xi = H/{(4\pi
F_\pi)^2}$. Thus, the domain of ChPT validity in a magnetic field
is $M^2_\pi/{(4\pi F_\pi)^2} \ll \xi < 1$. In the chiral limit,
the axial constant $F_\pi (M_\pi \rightarrow 0) \rightarrow const
\approx$ 80 MeV; hence, we have $0 < \xi < 1$.} $H\ll \mu^2_{hadr}
\sim (4\pi F_\pi)^2$ (electric charge $e$ is absorbed in $H$)
characteristic momenta in loops are small and QCD is effectively
described by chiral lagrangian $L_{eff}$ [2] which can be
represented as an expansion in powers of the external momenta
(derivatives) and the quark masses \be \label{L}
L_{eff}=L^{(2)}+L^{(4)}+L^{(6)}+\cdots \ee The leading term
$L^{(2)}$ in ($\ref{L}$) is similar to the Lagrangian of nonlinear
$\sigma$-model in external field  $V_\mu$ \be \label{CL2}
L^{(2)}=\frac{F^2_\pi}{4} {\rm Tr} (\nabla_\mu U^+ \nabla_\mu U)+
\Sigma {\rm Re Tr } ({\cal M} U^+ ), \ee
$$
\nabla_\mu U = \partial_\mu U - i \left [ U, V_\mu \right ] .
$$
Here $U$ stands for an unitary $SU(2)$ matrix (for two flavors),
$F_\pi=93$~Mev is the pion decay constant and parameter $\Sigma$
has the meaning of the quark condensate $\Sigma= |\langle \bar {u}
u \rangle |=|\langle \bar {d} d \rangle |.$ The external Abelian
magnetic field $H$, aligned along $z$-axis, corresponds to $V_\mu
(x) = (\tau^3 /2) A_\mu  (x)$, where vector potential $A_\mu$ may
be chosen as $A_1 (x) =-H x_2$. The difference of light quark masses
$m_u -m_d$ enters in the effective Lagrangian ($\ref{L}$) only
quadratically. To calculate the quantities which we are interested
in the chiral limit, we can put $m_u = m_d =0$ from the very
beginning. It means that we can take mass matrix to be diagonal
${\cal M}=m {\hat I}$.

In one-loop approximation of ChPT
the behavior of quark condensate in the presence of magnetic field and
finite temperature  was found in [8]. It proved to be enough to expand
$L_{eff}$ only up to quadratical terms of pion fields, i.e. neglect pion
interactions.  The result for the quark condensate is
[8]
$$ \frac{\Sigma (T,H)}{\Sigma}=
 1-\frac{T^2}{24 F^2_\pi}+\frac{H}{(4\pi
 F_\pi)^2} \ln 2-\frac{H}{2\pi^2F^2_\pi} \varphi
\left (\frac{\sqrt{H}}{T}\right)
 $$
\be
\varphi(\lambda)=\sum^\infty_{n=0}\int^\infty_0\frac{dx}{\omega_n(x)
(\exp(\lambda\omega_n(x))-1)}, ~~ \omega_n(x)=\sqrt{x^2+2n+1}
\label{22} \ee
In the chiral limit the function $\varphi$ depends only on the ratio
$\lambda=\sqrt H /T$ and has the following asymptotic behavior
[8]
\be
\varphi (\lambda \gg 1)={\sqrt \frac{\pi}{2\lambda}} e^{-\lambda}
+O(e^{-\sqrt 3 \lambda}),
\ee
\be
\varphi (\lambda \ll 1)=\frac{\pi^2}{6 \lambda^2} +
\frac{7\pi}{24\lambda} +\frac{1}{4} \ln \lambda +C +
\frac{\zeta (3)}{48\pi^2}\lambda^2 +O(\lambda^4).
\ee
Here $C=1/4(\gamma -\ln 4\pi -1/6),$ $\gamma=0.577...$ - Euler constant
and $\zeta (3)=1.202...$ is Riemann zeta-function.

To check Gell-Mann - Oakes - Renner relation for $\pi^0$ meson we
need to know the expressions for $M^2_{\pi^0}(H,T)$ and
$F^2_{\pi^0}(H,T)$. Firstly, let us find  the shift of $\pi^0$ mass in
magnetic field at finite temperature in the leading order of ChPT.
In this order it is enough to expand $L^{(2)}$ up to 4-pion
vertices. Choosing Weinberg parameterization for matrix $U$ \be
\label{L2} U=\sigma + \frac{i \pi^a \tau^a}{F_\pi},\quad
\sigma^2+\frac{\vec \pi^2}{F^2_\pi}=1, \ee we get the next order
terms of lagrangian $L^{(2)}$ \be \Delta L^{(2)}
=\frac{1}{2F_\pi^2}[\pi^0
\partial_\mu \pi^0 + \partial_\mu (\pi^+ \pi^-)]^2
-\frac{M^2_\pi}{8F^2_\pi} [2 \pi^+ \pi^- + (\pi^0)^2] \ee
Diagrams, contributing to shift of $M^2_{\pi^0}(H,T)$ are
represented in Fig. 1.

\unitlength=1mm
\begin{figure}[!tbh]~~~~~~~~~~~~~~~~~~~
\begin{picture}(30,30)
\put(0,0){\includegraphics{figs.1}}
\fmfLrt{-1.49113}{8.50887}{$\pi ^0$}%
\fmfLlt{31.49113}{8.50887}{$\pi ^0$}%
\fmfLlb{21.74121}{20.24129}{$\pi ^0$}%
\end{picture}~~~~~~~~~~~~~~~~~~~~~~~~~
\begin{picture}(30,30)
\put(0,0){\includegraphics{figs.2}}
\fmfLrt{-1.49113}{8.50887}{$\pi ^0$}%
\fmfLlt{31.49113}{8.50887}{$\pi ^0$}%
\fmfLlb{21.74121}{20.24129}{$\pi ^{+},\pi ^{-}$}%
\end{picture}
\caption{}
\label{fig.1}
\end{figure}
\unitlength=1pt

Using vertices from (7),
$M^2_{\pi^0}(H,T)$ can be represented as
\be
\label{M}
\frac{M^2_{\pi^0}(H,T)}{M^2_{\pi^0}}=1-\frac{1}{2F^2_{\pi^0}}
D^R_{T} (0) + \frac{1}{F^2_{\pi^0}}  D^R_{T,H}(0),
\ee
where $D^R_{T} (0)$ and $D^R_{T,H}(0)$ are propagators of scalar particles
at coinciding initial and final points. Index $"R"$ means that we
have subtracted from propagator its value at $H=0$, $T=0$.
The second term in
(8)
corresponds to Fig.1 when the particles
running in the loop are $\pi^0$-mesons noninteracting with magnetic field.
Temperature propagator $D^R_T(0)$ can be represented in the form
\be
\label{DT}
D^R_T (0)=D_T (x,x)- D_0 (x,x)=T \sum_{n=-\infty}^{+\infty}
\int \frac{d^3 {\bf p}}{(2\pi)^3} \frac{1}{\omega_n^2 +\omega^2 ({\bf p})}
-\int \frac{d^4 p}{(2\pi)^4}\frac{1}{p^2+ M^2_\pi},
\ee
where $\omega_n =2\pi n/\beta$, $\beta=1/T$,
$\omega^2 ({\bf p})= {\bf p^2} + M^2_\pi$.
Summing over Matsubara frequencies
\footnote{The following formula was used for summation (see, e.g. $\cite{Kap}$)
$$
\sum_{n=-\infty}^{+\infty}  \frac{1}{(2\pi n/\beta)^2 + \omega^2 } =
\frac{\beta}{2\omega} + \frac{\beta}{\omega (e^{\beta\omega}-1)}
$$}
in ($\ref{DT}$) we obtain (in the chiral limit $M^2_\pi \rightarrow 0$)
\be
\label{TT}
D^R_T (0)=\int \frac{d^3 {\bf p}}{(2\pi)^3}
\frac{1}{\omega (e^{\beta\omega}-1)}=
\frac{T^2}{12}
\ee

The third term in Eq.(8) corresponds to Fig.1 with charged
$\pi^{\pm}$-mesons
 in the loop and $D^R_{T,H} (0)$ denotes temperature propagator
(at coinciding points) of scalar particle moving in constant
magnetic field \be \label{DTH} D^R_{T,H} (0)=\frac{H}{2\pi}
\sum_{k=0}^\infty T \sum_{n=-\infty}^{+\infty}
\int_{-\infty}^{+\infty}\frac{dp_z}{2\pi} \frac{1}{\omega^2_n  +
\omega_H^2 (k,p_z)} - \int \frac{d^4 p}{(2\pi)^4}\frac{1}{p^2+
M^2_\pi} \ee and $\omega_H^2 (k,p_z)=p_z^2+M^2_\pi + H(2k+1)$ are
Landau levels. The first term in ($\ref{DTH}$) reproduces the
summation over eigenvalues of Hamiltonian of charged scalar
particle in magnetic field and  the degeneracy multiplicity of
$H/2\pi$ has been  taken into account for the Landau levels. It is
convenient to split $D^R_{T,H} (0)$ into two parts: "vacuum"
($T=0$) and "matter" ($T\neq 0$) pieces
\be D^R_{T,H} (0)= D^R_{0,H} (0)
+\tilde D_{T,H} (0), \ee where (in the chiral limit $M^2_\pi
\rightarrow 0$)
$$
D^R_{0,H} (0) = \frac{H}{8\pi^2}
\sum_{k=0}^\infty \int_{-\infty}^{+\infty}\frac{dp_z}{\omega_H (k,p_z)}-
\int \frac{d^4 p}{(2\pi)^4}\frac{1}{p^2}=- \frac{H}{16\pi^2}\ln 2
$$
\be
\tilde D_{T,H} (0) = \frac{H}{4\pi^2}\sum_{k=0}^\infty
\int_{-\infty}^{+\infty}\frac{dp_z}{\omega_H (e^{\beta \omega_H}-1)}=
\frac{H}{2\pi^2}\varphi \left(\frac{\sqrt H }{T}\right)
\label{TH}
\ee
Substituting ($\ref{TT}$) and ($\ref{TH}$) in ($\ref{M}$) we find one-loop
correction to $M^2_{\pi^0}$
\be
\label{MHT}
\frac{M^2_{\pi^0}(H,T)}{M^2_{\pi^0}}=1-\frac{T^2}{24F^2_\pi}
- \frac{H}{16\pi^2}\ln 2 + \frac{H}{2\pi^2}\varphi
\left(\frac{\sqrt H }{T}\right)
\ee

\unitlength=1mm
\begin{figure}[!tbh]~~~~~~~~~~~~~~~~~~~~~~~~~~~~~~~~~~~~~~~~~~~~~~~~
\begin{picture}(30,30)
\put(0,0){\includegraphics{figs.3}}
\fmfLrt{-1.49113}{8.50887}{$A_{\mu }$}%
\fmfLlt{31.49113}{8.50887}{$A_{\mu }$}%
\fmfLlb{21.74121}{20.24129}{$\pi ^{+},\pi ^{-}$}%
\end{picture}
\caption{}
\label{fig.2}
\end{figure}
\unitlength=1pt

To calculate renormalization of axial coupling constant $F_{\pi^0}$
we consider the correlator of two axial currents
\be
\Pi^A_{\mu\nu}=i\int d^4 x e^{iqx} \langle A_\mu (x) A_\nu (0)\rangle,
\ee
where $A_\mu  =\bar q \gamma_\mu \gamma_5 (\tau^3 /2) q$
is the third component of axial current.
As $F_\pi$ and its shift are non-zero in the chiral limit, we put
$M_{\pi}^2=0$ from the very beginning.
Then this correlator becomes transverse
ant takes the form
\be
\Pi^A_{\mu\nu} = F^2_\pi (g_{\mu\nu}-\frac{q_\mu q_\nu}{q^2})
\ee
and we will calculate the coefficient before $g_{\mu\nu}$.
In the first order of ChPT the relevant part of chiral lagrangian describing
the interaction of pions with sources of axial current $a_\mu$ is
\be
\Delta L_A =\frac12 F_\pi^2 a_\mu^2 - a_\mu^2 \pi^+ \pi^-
\ee
The shift of $F_{\pi^0}$ is given by diagram in Fig. 2 and can be written
in the form
\be
\frac{F^2_{\pi^0}(H,T)}{F^2_{\pi^0}}=1-
\frac{2}{F^2_{\pi^0}}  D^R_{T,H}(0),
\ee
Using relations for $ D^R_{T,H}$ we finally obtain  correction to axial
coupling constant  in the magnetic field $H$ at finite temperature $T$.
\be
\label{FHT}
\frac{F^2_{\pi^0}(H,T)}{F^2_{\pi^0}}=1+
 \frac{H}{8\pi^2}\ln 2 - \frac{H}{\pi^2}\varphi
\left(\frac{\sqrt H }{T}\right)
\ee

Making use of ($\ref{MHT}$), ($\ref{FHT}$) and ($\ref{22}$) we find the following equation
\be
F^2_{\pi^0}(T,H) M^2_{\pi^0}(T,H) =2m\Sigma (T,H),
\ee
which means that in the first order of ChPT
Gell-Mann - Oakes - Renner relation remains valid in
the magnetic field at finite temperature. Hence, the scheme
of "soft" breaking of chiral symmetry by quark masses is not affected
by magnetic field and temperature.

Let us consider different limiting cases.
When magnetic field is switched off we reproduce well-known results
at finite temperature [6]
\be
\frac{\Sigma(T)}{\Sigma}=1-\frac{T^2}{8F_\pi^2},~~~~
\frac{M^2_\pi (T)}{M^2_\pi}=1+\frac{T^2}{24F_\pi^2},~~~~
\frac{F^2_\pi (T)}{F^2_\pi}=1-\frac{T^2}{6F_\pi^2}.
 \ee
If we set $T=0$ then we get the results obtained in [7]
\be
\frac{\Sigma(H)}{\Sigma}=1+\frac{H\ln 2}{(4\pi F_\pi)^2},~~~
\frac{M^2_\pi (H)}{M^2_\pi}=1-\frac{H\ln 2}{(4\pi F_\pi)^2},~~~
\frac{F^2_\pi (H)}{F^2_\pi}=1+\frac{2H\ln 2}{(4\pi F_\pi)^2}.
 \ee
In both limiting cases: $\sqrt H/T \ll 1$ and $\sqrt H/T \gg 1$
all corrections are governed by $\varphi$ and can be easily
found from (4) and (5), and
for the quark condensate these corrections were
obtained in [8].

It has been shown in [8] that the quark
condensate is "frozen" by the magnetic field when both temperature
$T$ and magnetic field $H$ are increased according to the
 $H=const \cdot T^2$ law. The same effect can be observed for
$M_{\pi^0} (H,T)$ and $F_{\pi^0} (H,T)$ behavior. To find a regime
of $\pi^0$ mass "freezing" we should solve the equation followed from
(14)
\be
1+\frac{3}{2\pi^2}\lambda^2 \ln 2 -\frac{12}{\pi^2}\lambda^2
\varphi (\lambda)=0
\ee
Numerically, $\lambda_M \approx 0.069...$. Hence,  $\pi^0$ mass remains
unchanged when $H=\lambda_M T^2$. Similar consideration can be done
for $F_{\pi^0} (H,T)$. In this case appropriate equation is
\be
\varphi (\lambda) = \frac18 \ln 2
\ee
and $\lambda_F \approx 2.323...$

3. In the present paper we found the behavior of $F_{\pi^0}$ and $M_{\pi^0}$
constants in a magnetic field at finite temperature. It was demonstrated
that for $F_{\pi^0}$ and $M_{\pi^0}$
effect of "freezing" takes place when both temperature
$T$ and magnetic field $H$ increase according to the
$H=const \cdot T^2$ law and two appropriate constants $\lambda_F$
and $\lambda_M$ were numerically found.
It was shown that
Gell-Mann - Oakes - Renner relation for $\pi^0$ meson remains valid,
i.e. the mechanism of "soft" breaking of chiral symmetry by light quark
masses is not affected by magnetic field and temperature.

N.A. acknowledges the financial support of RFBR grant 00-02-17836
and INTAS grant CALL 2000 No.110. I.S. acknowledges the financial support of
 CRDF  No. RP2-2247,
RFBR No. 00-02-17808, and grant INTAS 2000-587.

 \end{document}